\documentclass[nofootinbib,amsmath,amssymb,showpacs,showkeys,aps,reprint]{revtex4-1}
\usepackage[utf8,latin1]{inputenc}
\usepackage{graphicx}
\usepackage{dcolumn}
\usepackage[dvipsnames]{xcolor}
\usepackage[T1]{fontenc}

\usepackage{mathrsfs}  
\usepackage{cases}
\usepackage{bm}
\usepackage{academicons}
\usepackage{mathtools, nccmath}
\usepackage{fancyhdr}
\usepackage{tikz,xcolor}
\newcommand{\qm}[1]{``#1''}
\usepackage{tensor}
\usepackage[normalem]{ulem}
\usepackage{lipsum}
\usepackage{soul}
\usepackage{cancel}
\usepackage{stackengine,scalerel}
\usepackage{mathabx}
\usepackage{hyperref}
\usepackage{tabularx}
\hypersetup{colorlinks, linkcolor={blue},citecolor={blue},urlcolor={blue}}

\definecolor{lime}{HTML}{A6CE39}
\DeclareRobustCommand{\orcidicon}{
	\begin{tikzpicture}
	\draw[lime, fill=lime] (0,0) 
	circle [radius=0.16] 
	node[white] {{\fontfamily{qag}\selectfont \tiny ID}};
	\draw[white, fill=white] (-0.0625,0.095) 
	circle [radius=0.007];
	\end{tikzpicture}
	\hspace{-2mm}
}

\fancypagestyle{plain}{%
  \fancyhf{}
  \fancyfoot[C]{\iffloatpage{}{\thepage}}
  }
\foreach \x in {A, ..., Z}{%
	\expandafter\xdef\csname orcid\x\endcsname{\noexpand\href{https://orcid.org/\csname orcidauthor\x\endcsname}{\noexpand\orcidicon}}
}

\begin{document}

\title[Shadow signatures and energy accumulation in  Lorentzian-Euclidean black holes]{Shadow signatures and energy accumulation in  Lorentzian-Euclidean black holes}

\author{Emmanuele Battista\orcidA{}$^{1}$\vspace{0.5cm}}\email{ebattista@lnf.infn.it} 
\author{Salvatore Capozziello\orcidB{}$^{2,3,4}$\vspace{0.5cm}} \email{capozziello@na.infn.it}
\author{Che-Yu Chen\orcidC{}$^{5}$\vspace{0.5cm}} \email{b97202056@gmail.com}

\affiliation{$^1$ Istituto Nazionale di Fisica Nucleare, Laboratori Nazionali di Frascati, 00044 Frascati, Italy,\\
$^2$ Dipartimento di Fisica ``Ettore Pancini'', Complesso Universitario 
di Monte S. Angelo, Universit\`a degli Studi di Napoli ``Federico II'', Via Cinthia Edificio 6, 80126 Napoli, Italy,\\
$^3$ Istituto Nazionale di Fisica Nucleare, Sezione di Napoli, Complesso Universitario 
di Monte S. Angelo, Via Cinthia Edificio 6, 80126 Napoli, Italy,\\
$^4$  Scuola Superiore Meridionale, Largo San Marcellino 10, 80138 Napoli, Italy, \\
$^5$ RIKEN iTHEMS, Wako, Saitama 351-0198, Japan
}

\date{\today} 

\begin{abstract}

The Lorentzian-Euclidean black hole has been recently introduced as a geodesically complete spacetime featuring a signature shift at the event horizon where causal geodesics are precluded from reaching the central $r=0$ singularity. In this paper, we investigate the shadows produced by this geometry to identify deviations from the standard Schwarzschild solution. Our analysis reveals an excess intensity in the inner shadow region that points to a potential observational signature of the novel behavior of light rays propagating near the event horizon. This excess could be a probe for horizon-scale modifications of black hole geometries. Furthermore, although the horizon surface of the Lorentzian-Euclidean black hole continuously accumulates photons and energy, we show that its backreaction response differs from that of stable light rings found in various exotic compact objects.

\end{abstract}

\maketitle

\section{Introduction}

The recent observations of gravitational waves from inspiralling compact binaries by the LIGO-Virgo-KAGRA network \cite{LIGOScientific2025}, together with the remarkable images of the supermassive black holes in M87 and the Milky Way released by the Event Horizon Telescope (EHT) collaboration \cite{EventHorizonTelescope:2019dse,EventHorizonTelescope:2022wkp}, have substantially boosted interest in black hole physics. These breakthroughs have intensified the exploration of strong-gravity and horizon-scale phenomenology \cite{Bambi2017,Barack2018,Cardoso2019, Molla:2025yoh, Capozziello:2023tbo}, as well as effective and quantum-gravity-inspired modifications of black hole spacetimes \cite{Calmet2019,Donoghue2022,Chen:2023wdg,Battista2023-quantum,Zhang2023,DelPiano2023,DelPiano2024,Shu2024,Hohenegger2025},  including proposals for regular models. Such scenarios range from further developments and extensions of the well-known Bardeen \cite{Singh2023,Zeng2025,Ma2025,Vertogradov2025b,Capozziello:2025ycu}, Dymnikova \cite{Konoplya2024,Alshammari2025,Vertogradov2025}, and Hayward \cite{Fathi2021,Gohain2024,Bora2025,Waseem2025,Liang2025,Naseer2025,Boos2025} metrics, to constructions supported by nonlinear electrodynamics sources \cite{Culetu2014,Miao2024,Sultan2025,Junior2025,Huang2025,Contreras2025}, to black-bounce geometries \cite{Simpson2018,Ditta2024}, and to frameworks incorporating quantum-gravity corrections of the near-horizon geometry \cite{Xiang2013,Brahma:2020eos,Bhandari2024,Belfaqih:2024vfk,Calza2024a,Calza2024b,Calza2025c,Calza2025d,Li-S2025a}.

Recently, a new paradigm, dubbed Lorentzian-Euclidean black hole, has been set forth \cite{Capozziello:2024ucm,Capozziello:2025wwl,DeBianchi2025}. In this setup, a change in the metric signature occurs, replacing the usual Lorentzian structure with an ultrahyperbolic one beyond the event horizon, while causal geodesics are hindered from reaching the central $r=0$ singularity, which is thus effectively \emph{avoided}. 

Motivated by the fact that, in general, the theoretical appeal of regularized spacetimes stems not only from their ability to evade curvature singularities, but also from the rich phenomenology they entail, in this paper we will analyze the shadow images cast by the Lorentzian-Euclidean black hole in order to reveal potential observational signatures. Since one relevant feature of this geometry is that (massive particles and) light rays require an infinite affine parameter to reach the event horizon \cite{Capozziello:2024ucm,Capozziello:2025wwl,DeBianchi2025}, we will identify the image signatures that can capture this nontrivial behavior. 

Furthermore, the accumulation of photons and energies at $r=2M$, i.e., the event horizon, raises the question of whether the ensuing backreaction could potentially induce spacetime instabilities, similar to those associated with stable light rings in various models of horizonless compact objects \cite{Keir:2014oka,Cardoso:2014sna,Cunha:2022gde}. We will address this issue by highlighting a substantial difference between such stable light rings and the event horizon of  Lorentzian-Euclidean black holes.

The outline of the paper is as follows. In Sec.~\ref{Sec:II}, we briefly review the Lorentzian-Euclidean black hole and examine the dynamics of light rays. In Sec.~\ref{Sec:shadow}, we investigate the shadow images and identify the potential observational signatures. Then, in Sec.~\ref{Sec:lightring}, we discuss how the aggregation of energies near the event horizon may affect the underlying geometry. Finally, we draw our conclusions in Sec.~\ref{Sec:Conclusion}. 

Throughout the paper, we use units $G=c=1$.

\section{Lorentzian-Euclidean black holes and photon dynamics}\label{Sec:II}

In  the Schwarzschild coordinates $\{t,r,\theta,\phi\}$, the Lorentzian-Euclidean Schwarzschild metric takes the form \cite{Capozziello:2024ucm,Capozziello:2025wwl,DeBianchi2025}
\begin{align}
    d s^2 =& g_{\mu \nu} d x^\mu d x^\nu=- \varepsilon \left( 1-\frac{2M}{r} \right)  d t^2  
+ \dfrac{d r^2 }{\left(1-\frac{2M}{r}\right)} 
\nonumber \\
&+ r^2 \left(d \theta^2  + \sin^2 \theta \; d \phi^2 \right)\,,
\label{Lorentzian-Euclidean-Schwarzschild}
\end{align}
where $M$ is the black hole mass and 
\begin{align}
\varepsilon = {\rm sign} \left( 1 - \frac{2M}{r}\right)= 2 H \left( 1 - \frac{2M}{r}\right)-1\,,
\label{epsilon-of-r}
\end{align}
with the step function $H\left(1-2M/r\right)$  normalized in such a way that $H(0)=1/2$. 

The metric becomes degenerate at the event horizon $r=2M$, where its determinant goes to zero and the signature changes from Lorentzian ($r>2M$) to ultrahyperbolic  ($r<2M$)\footnote{Degenerate extensions of the Schwarzschild geometry have also been considered in Refs. \cite{Kaul2016,Kaul2017}, while applications to gravitational collapse of the signature-changing mechanism have been recently studied in Ref. \cite{Bahn2025}.}. In the inner region ($r<2M$), the spacetime exhibits features similar to the  Euclidean Schwarzschild geometry (see e.g. Refs.  \cite{GH1977,Esposito1992,Battista-Esposito2022,Garnier2024,Garnier2025} for further details). This behavior can be understood in terms of a time coordinate $t$ that becomes imaginary upon crossing $r=2M$, a phenomenon we proposed to interpret via the notion of \emph{atemporality} \cite{DeBianchi2025}.

We begin this section by reviewing the main facet of the Lorentzian-Euclidean black holes. First,  the signature shift generates  Dirac-delta-like factors in both the Ricci and Weyl parts of the Riemann tensor, which are \emph{a priori} ill-defined and thus require an appropriate regularization scheme,   outlined in Sec. \ref{Sec:regul-scheme}. Second, a key aspect of the model is that no causal geodesic passes through the event horizon. This feature modifies the causal structure of the spacetime, as we briefly review in Sec.  \ref{Sec:causal-structure}. After these premises, we conclude the section by investigating equatorial photon motion in Sec. \ref{Sec:geodesic}, where the obtained results set the stage for the shadow analysis in Sec. \ref{Sec:shadow}.

\subsection{The regularization scheme} \label{Sec:regul-scheme}

In general relativity, junction conditions provide a systematic framework for consistently matching two metrics across a hypersurface $\Sigma$ that divides the spacetime into two regions \cite{Visser1995-book,Barrabes-book2003,Poisson2009-book}. The Einstein equations are then understood in a distributional sense, and the standard treatment follows the  Darmois-Israel formalism. This pattern relies on two requirements whose purpose is to eliminate, or at least give a physical meaning to, singular distribution-valued contributions arising in the geometric quantities constructed from the metric.

The first matching relation is designed to ensure that the Christoffel symbols are well defined as distributions, and demands that the induced three-metric on $\Sigma$ be continuous. The second junction condition is imposed to prevent $\delta$-function terms in the Riemann tensor by enforcing continuity of the extrinsic curvature across $\Sigma$. When  this request fails, the resulting  Dirac-delta-like
singularities in the Einstein tensor are interpreted physically as a thin (delta-function) layer of matter supported on $\Sigma$, which thus carries a distributional surface stress-energy tensor.

When $\Sigma$ is timelike or spacelike, singular components are confined to the Ricci tensor; in contrast, if $\Sigma$ is null, Dirac-delta factors may also appear in the Weyl tensor and correspond to a gravitational shock wave.

In the Lorentzian-Euclidean setup,    $\Sigma$ is the null hypersurface identified with the event horizon, and it now has the additional property of defining the boundary where the metric becomes degenerate and undergoes a signature change.  Furthermore,  the curvature tensor develops  factors of the form $\delta\left(r-2M\right)/\varepsilon$ and $\delta^2\left(r-2M\right)$, which are \emph{formally} ill-defined in the  Schwartz theory of distributions \cite{Lieb2001}. These features go beyond the usual thin-shell setting, as they reflect the rich and intricate structure underlying geometry \eqref{Lorentzian-Euclidean-Schwarzschild}. As a consequence, the Darmois-Israel theory can no longer be applied in a straightforward way, and thus we need an alternative method, which however has the same aim: to define all geometric objects consistently as distributions across   (the now signature-changing) $\Sigma$. 

For this reason, in Ref. \cite{Capozziello:2024ucm}, we have devised a regularization scheme based on three ingredients. First, we introduce the smooth family of functions 
\begin{equation}
\varepsilon(r)=\frac{\left(r-2M\right)^{1/(2\kappa+1)}}{\left[(r-2M)^2+\rho\right]^{1/2(2\kappa+1)}}\,,\label{epsilonr}
\end{equation}
which approximates  the sharp profile of the  sign function \eqref{epsilon-of-r}  for $\rho/M^2 \to 0^+$ and  large values of the positive integer  $\kappa $ (although the regularization setup only demands $\kappa \geq 1$ \cite{Capozziello:2024ucm}). In our regularization process, to preserve the Dirac-delta content of curvature tensors we first compute them for the smooth metrics built from Eq. \eqref{epsilonr} and only afterward take $\rho/M^2 \to 0^+$; in this limit, the derivatives $\varepsilon^\prime$ and $\varepsilon^{\prime \prime}$  give rise to the Dirac delta and its first-order derivative, respectively (hereafter, the prime denotes differentiation with respect to the radial variable $r$). In this way,  combinations of the type $\delta\left(r-2M\right)/\varepsilon$ become proportional to $\delta\left( r-2M \right)/\vert r-2M \vert^n$, with  $n>0$. Interestingly,  terms of this kind also appear in the post-Newtonian treatment of inspiralling compact binaries in gravitational-wave theory \cite{Blanchet2014-review,Poisson-Will2014}, where they are handled using the Hadamard \emph{partie finie} prescription. The second feature of our approach is precisely to adopt this technique, which 
consistently assigns zero (in the finite-part sense) to expressions of the form $\delta(r-2M)/\vert r-2M\vert$ when interpreted distributionally  \cite{Poisson-Will2014,Blanchet2000}. The third ingredient addresses curvature tensor components involving $\delta^{2}(r-2M)$. Also in this case, the gravitational-wave literature, in particular the framework of impulsive (shock) waves, provides a useful guideline. Indeed, in these models the problem is solved by a simple but effective method, as one can argue that quadratic-in-delta quantities do not contribute to the distributional curvature if they are multiplied by coefficients that vanish on the support of $\delta$, i.e., at the hypersurface where the impulse is located  \cite{Dray1984,Sfetsos1995,Battista_Riemann_boosted,Battista2015-proc}. In our setup, this translates into the fact that products such as $f(r) \delta^2\left(r-2M\right)$  are zero in the distributional sense whenever $f(2M)=0$.

Remarkably, our regularization procedure shows that all pathological $\delta$-contributions have vanishing distributional action. Specifically, the regularized Riemann and Weyl tensors depend only on $\varepsilon$,  while the regularized Ricci tensor and scalar curvature are identically zero (further details, including explicit calculations and cross-checks of our conclusions, can be found in Ref. \cite{Capozziello:2024ucm}). This shows that the  Einstein tensor is well defined as a distribution and is zero in the distributional sense as $\rho/M^2\to0^+$, for any $\kappa \geq 1$ fixed. In particular, all would-be $\delta$-supported terms on $\Sigma$ are distributionally vanishing. Consequently, the metric \eqref{Lorentzian-Euclidean-Schwarzschild} can be interpreted as a valid signature-changing solution of the vacuum Einstein equations in the distributional sense. The ensuing geometry contains no surface layers or impulsive waves at the horizon. Moreover, the regularized Kretschmann scalar retains its usual form, i.e.,  $\mathcal{K} = 48 M^2/r^6$, indicating that the only true singularity of the Lorentzian-Euclidean spacetime resides at $r=0$. 

A final important remark is in order. The regularization procedure discussed above is an \emph{intrinsic} feature of our framework. A continuous and differentiable $\varepsilon(r)$ of the form \eqref{epsilonr} is essential for regularizing the geometry, as otherwise the spacetime structure becomes ill-defined and key quantities like the Riemann tensor acquire distributional singularities. Consequently, the observational signatures discussed below are not artifacts of employing a smoothed step function, but constitute genuine predictions of the model. 

\subsection{Causal structure and the horizon }\label{Sec:causal-structure}

Although the regularized curvature invariants remain formally divergent at $r=0$,  the analysis of timelike and null radial geodesics  (as well as of radially accelerated paths) reveals that their velocity vanishes at the event horizon and becomes imaginary inside it \cite{Capozziello:2024ucm,Capozziello:2025wwl}. Therefore, no causal trajectory traverses the event horizon and infalling particles are prevented from reaching the center of the black hole, the singularity being thereby effectively \emph{avoided}. Furthermore, the spacetime turns out to be geodesically complete, as the affine parameter becomes infinite upon approaching $r=2M$.  

These distinctive behaviors leave a clear imprint on the global causal structure of the Lorentzian-Euclidean spacetime, which we have analyzed in detail in Ref. \cite{Capozziello:2025wwl}. Indeed, the corresponding Penrose diagram consists only of the two asymptotically flat outer Lorentzian regions (namely, the right-hand and left-hand exteriors of the maximally extended spacetime) bounded by the future and past event horizons. The reason is that the degenerate surface $\Sigma$ acts as a two-way causal barrier, rather than the standard one-way membrane of the Schwarzschild metric, and the entire inner sector, including the singularity, is causally disconnected from the exterior in \emph{both} directions. This suggests that the Lorentzian-Euclidean geometry gives rise to a new horizon-bearing configuration that extends the usual general-relativistic black hole construction.  Nevertheless, our model retains key black hole properties. Mirroring the Schwarzschild case,  $\Sigma$ is a two(-plus-one)-dimensional null surface with cross-sectional area  $4\pi(2M)^2$. Although the underlying mechanism differs from Einstein gravity, it can still be regarded as a surface of no return, defined, as usual, as the boundary (of the closure) of the causal past of future null infinity. Above all, the horizon continues to render the internal region, which is the Euclidean domain in our setup,  invisible to the outside. Finally,  an external observer at infinity never sees a particle crossing the horizon, exactly as in the Schwarzschild solution.

Therefore, the Lorentzian-Euclidean model is fully adequate for the present analysis, since the geodesic and optical observables discussed here are entirely determined by the outer Lorentzian zone.

We can thus conclude that the Lorentzian-Euclidean geometry is able to reproduce many black hole features globally, while exhibiting characteristic near-horizon departures from the ordinary framework. The investigation developed below will further strengthen this statement.

\subsection{Photon geodesics}\label{Sec:geodesic}

The investigation of the shadow images generated by the Lorentzian-Euclidean black holes, addressed in Sec. \ref{Sec:shadow}, requires extending the study of photon motion beyond the radial case. For this reason, we now consider generic null geodesics and exploit the spherical symmetry of the geometry \eqref{Lorentzian-Euclidean-Schwarzschild} to restrict our attention to orbits on the equatorial plane $\theta=\pi/2$. 

The normalization condition $g_{\mu\nu}\dot{x}^\mu\dot{x}^\nu=0$ yields 
\begin{equation}
-\varepsilon\left(1-\frac{2M}{r}\right)\dot{t}^2+\frac{\dot{r}^2}{1-2M/r}+r^2\dot{\phi}^2=0\,,\label{photonconstraint}
\end{equation}
where a dot denotes differentiation with respect to the affine parameter $\lambda$. The dynamics admits two conserved quantities, $\Omega$ and $\ell$, which represent the photon energy and angular momentum, respectively. While the latter takes the standard form
\begin{equation}
\ell=r^2\dot{\phi}\,,\label{lphoton}
\end{equation}
the former can be expressed as \cite{Capozziello:2025wwl}
\begin{equation}
\Omega=\mathcal{F}\varepsilon\left(1-\frac{2M}{r}\right)\dot{t}\,,\label{ephoton}
\end{equation}
where the function $\mathcal{F}$ complies with the identity $\mathcal{F}=\sqrt{\mathcal{F}^2}$ and is introduced to ensure the correct sign and physical meaning of $\Omega$ in both the Lorentzian ($r>2M$) and Euclidean ($r<2M$) domains.

Inserting Eqs.~\eqref{lphoton} and \eqref{ephoton} into the null constraint relation \eqref{photonconstraint}, we get
\begin{equation}
\dot{r}^2=\frac{\Omega^2}{\mathcal{F}^2\varepsilon}-\frac{\ell^2}{r^2}\left(1-\frac{2M}{r}\right)\,,\label{rdotequation}
\end{equation}
and then, by expressing Eq. \eqref{ephoton} as $\Omega=\alpha_p\mathcal{F}\varepsilon$ in terms of some positive-definite bounded function $\alpha_p$ \cite{Capozziello:2025wwl}, we  obtain
\begin{equation}
\sqrt{\varepsilon}d\phi=\pm\frac{dr}{r\sqrt{r^2/b^2-\left(1-2M/r\right)/\varepsilon}}\,,\label{angularint}
\end{equation}
where $b:= \ell/\alpha_p$  is the impact parameter of light rays, and the upper (resp. lower) sign corresponds to outgoing (resp. ingoing) motion. 

To integrate Eq.~\eqref{angularint}, we proceed as follows.  At the outset, we replace the sign function $\varepsilon$ on the right-hand side with the  regularizing family of smooth functions   \eqref{epsilonr}, and  then, drawing on the treatment of Ref.~\cite{Capozziello:2025wwl}, we recast the  term on the left-hand side  as 
\begin{equation}
\sqrt{\varepsilon}d\phi=d\left(\sqrt{\varepsilon}\phi\right)\,.\label{magictrick}
\end{equation}
This identity can be established either by a direct computation or via a distributional approach involving the Hadamard prescription (the procedure is analogous to that used in Eqs. (90)--(92) of Ref. \cite{Capozziello:2025wwl}).   In this way, we can first evaluate the right-hand side of Eq.~\eqref{angularint} to obtain $\Delta(\sqrt{\varepsilon}\phi)$ along a trajectory. Assuming the motion starts at $\phi=0$, we then divide the result by $\sqrt{\varepsilon}$ to properly recover the elapsed angular coordinate. This step is crucial for paths with impact parameters smaller than those of spherical photon orbits, as they can approach arbitrarily close to $r=2M$. After performing the integration of Eq.~\eqref{angularint}, the photon trajectories $r(\phi)$ for a given $b$ can be finally determined.

\begin{figure}[htbp!]
\centering
\includegraphics[width=230pt]{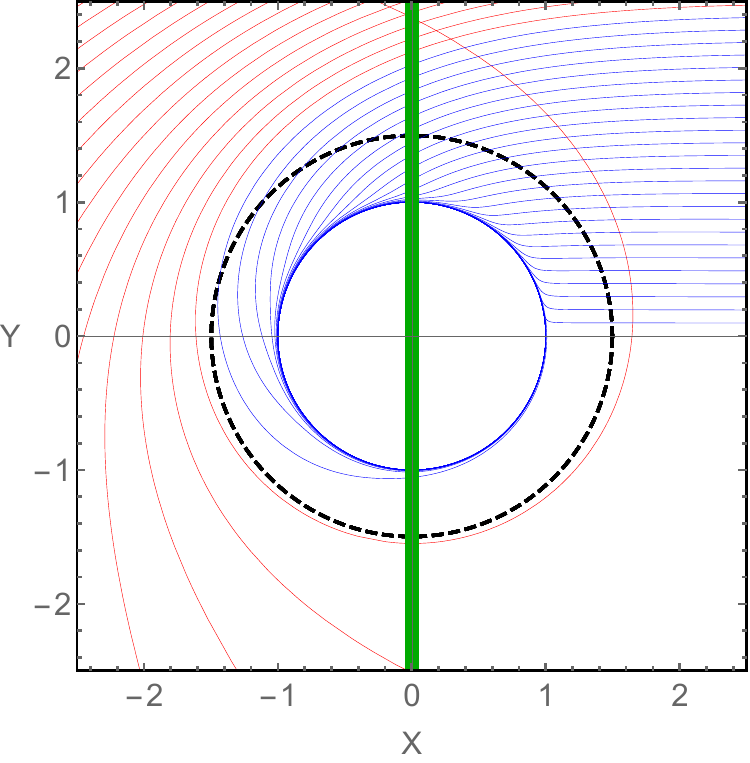}
\caption{Photon trajectories in the equatorial plane of the Lorentzian-Euclidean black hole. The black dashed circle marks the photon sphere  (see Eq. \eqref{photon-sphere-radius}), while the green vertical line denotes the thin accretion disk plane (see Sec.~\ref{Sec:shadow}). All orbits originate from the point $(X,Y)=(100,0)$; light rays with $b>b_c$ (resp. $b<b_c$), where $b_c$ is given in Eq. \eqref{impact-b-c},  are shown in red (resp. blue). We set $\kappa=1$,   $\rho=1/400$, and adopt  units  in which $2M=1$.}
\label{fig:trajectory}
\end{figure}
\begin{figure*}[htbp!]
\centering
\includegraphics[width=200pt]{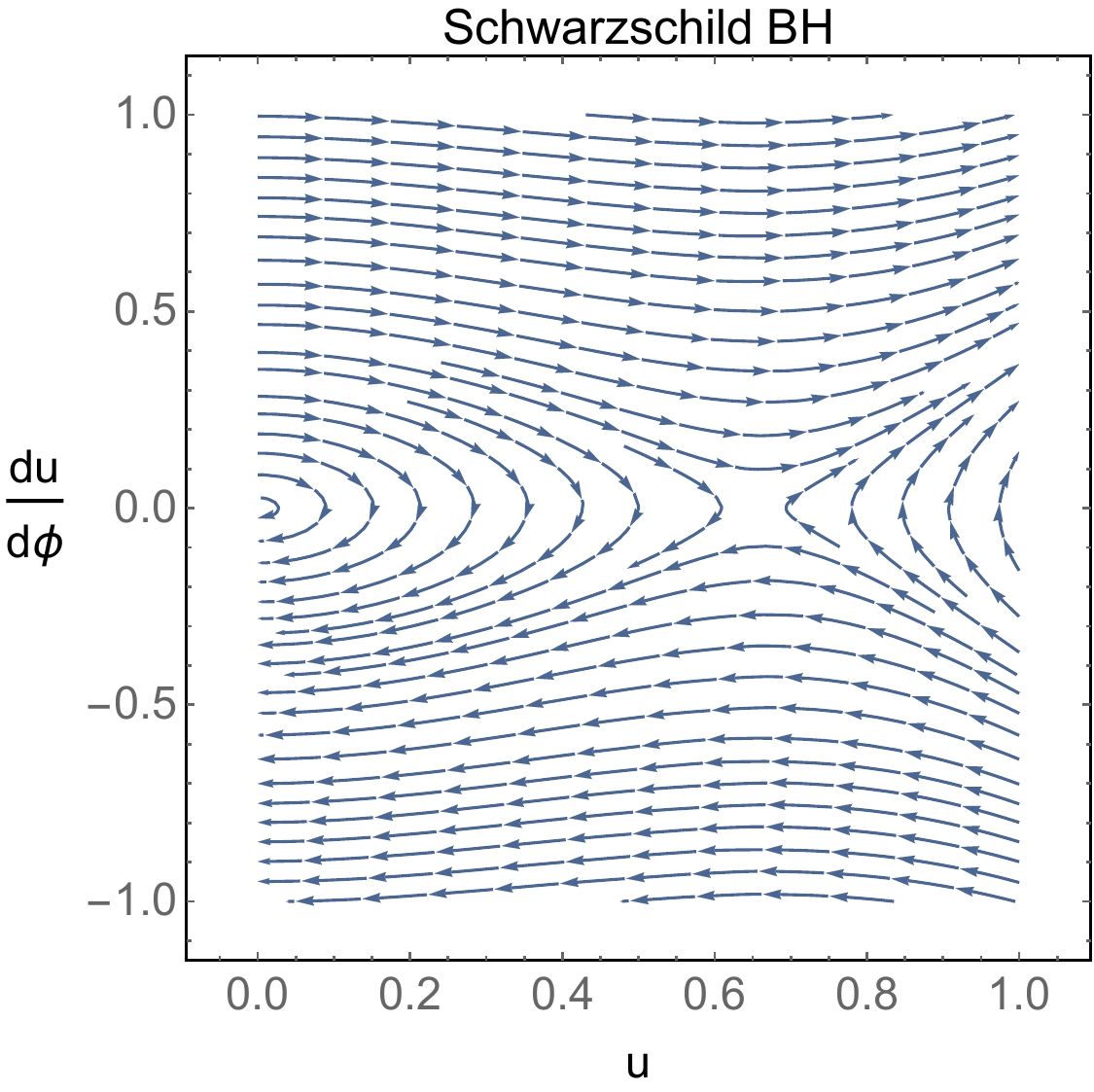}\qquad\qquad
\includegraphics[width=200pt]{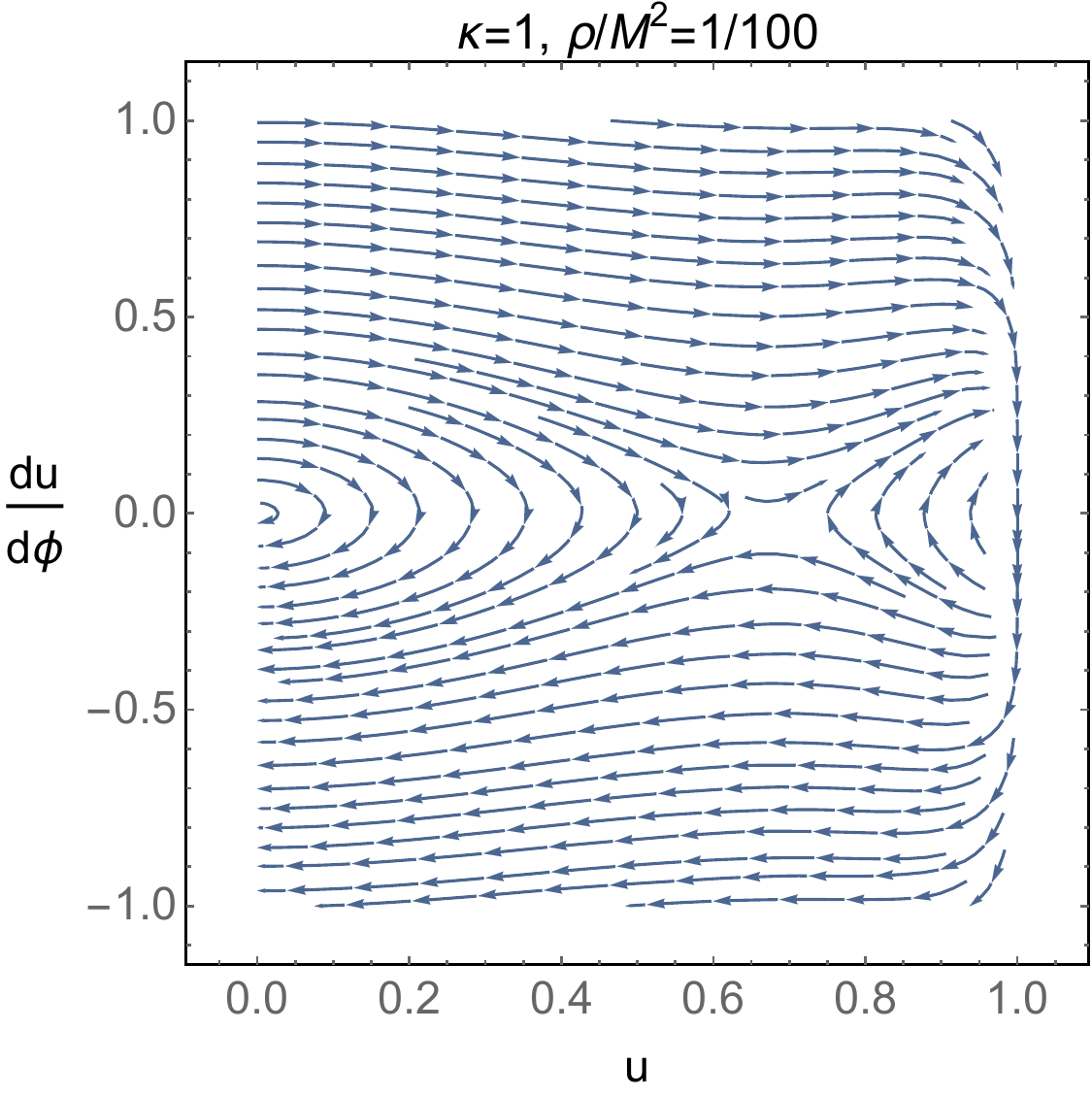}
\caption{Phase-space portraits of the photon dynamics in terms of $u\equiv 2M/r$ and $du/d\phi$ in the Schwarzschild spacetime (left) and the Lorentzian-Euclidean black hole spacetime (right) with $\kappa=1$ and $\rho/M^2=1/100$. For the Lorentzian-Euclidean black hole, the point $(u,du/d\phi)=(1,0)$ represents a degenerate attractor.}
\label{fig:phase}
\end{figure*}

The photon orbits around the Lorentzian-Euclidean black hole are displayed in Fig.~\ref{fig:trajectory}, where we have introduced the coordinates $X$ and $Y$ in the usual way: $X=r(\phi) \cos \phi$, $Y=r(\phi) \sin \phi$. The black dashed circle stands for the photon sphere, where light rays undergo unstable circular motion. These trajectories are determined by solving the equation 
\begin{align}
    \partial_r\left[\left(1-2M/r\right)/\varepsilon(r)r^2\right]=0\,, 
\end{align}
which differs from the  standard photon sphere equation $\partial_r(g_{tt}/r^2)=0$ because of the  definitions of the quantities $\Omega=\alpha_p\mathcal{F}\varepsilon$ and $\ell=b\alpha_p$ that shall be substituted in Eq.~\eqref{rdotequation}. Since the function $\varepsilon(r)$ given in Eq.~\eqref{epsilonr} rapidly approaches unity as one moves away from $r=2M$, the radius $r_\textrm{ph}$ of the photon sphere can be approximated to first order by
\begin{equation}
r_\textrm{ph}\approx3M\left[1-M\varepsilon'(3M)/2\right]\,.
\label{photon-sphere-radius}
\end{equation}
In view of the fact that  $\varepsilon'(r)>0$, in our setup, $r_\textrm{ph}$ is slightly smaller than the ordinary Schwarzschild value of $3M$. Furthermore, the impact parameter $b_c$ of the photon sphere, which represents the infinitely lensed ring on the image plane, can be obtained by solving 
\begin{align}
 r_\textrm{ph}^2/b_c^2-(1-2M/r_\textrm{ph})/\varepsilon(r_\textrm{ph})=0\,.  
\end{align}
Up to the first order, we get
\begin{equation} \label{impact-b-c}
b_c\approx\frac{\sqrt{27}M}{2}\left[1+\varepsilon(3M)\right]\,,
\end{equation}
which is also slightly smaller than the usual Schwarzschild value $\sqrt{27}M$ since $\varepsilon<1$. A similar trend, i.e., the shrinking of the photon sphere and its associated impact parameter, has been found in several nonsingular black hole models \cite{Dymnikova:1992ux,Hayward:2005gi,Nicolini:2005vd,Simpson:2019mud}. A common physical interpretation is that the new physics responsible for the singularity resolution effectively weakens the gravitational interaction \cite{Eichhorn:2022oma}. As a consequence, the photon sphere forms at a smaller radius, giving rise to a reduced shadow size. It is remarkable that the Lorentzian-Euclidean black hole, where the singularity is actually avoided, displays the same qualitative behavior.

The radius of the photon sphere $r_\textrm{ph}$ and the corresponding impact parameter $b_c$ essentially distinguish the light rays that are trapped by the black hole from those that are just scattered around. Orbits with $b>b_c$, illustrated by the red paths in Fig.~\ref{fig:trajectory}, have radial turning points outside the photon sphere and thus behave in a way closely analogous to the standard Schwarzschild case. On the other hand, photon trajectories having $b<b_c$, shown in blue in Fig.~\ref{fig:trajectory}, develop in a manner significantly different. Although they can reach the region in the vicinity of $r=2M$, they do not cross the event horizon but instead circle the black hole infinitely many times at radii extremely close to $r=2M$. This dynamical evolution aligns with the outcome of Ref.~\cite{Capozziello:2025wwl}, which demonstrates that photons take an infinite affine parameter to arrive at the event horizon of Lorentzian-Euclidean black holes. 

In Fig.~\ref{fig:phase}, we show the phase-space portraits of the photon dynamics in terms of $u\equiv 2M/r$ and $du/d\phi$ for the Schwarzschild black hole (left) and the Lorentzian-Euclidean black hole (right). In both cases, the photon sphere corresponds to a saddle point located at $u\approx 2/3$ and $du/d\phi=0$ (we notice that the stable fixed point at asymptotic infinity $(u,du/d\phi)=(0,0)$ is not relevant in our discussion). However, for the Lorentzian-Euclidean black hole,  ingoing trajectories are strongly bent toward a fixed point $(u,du/d\phi)=(1,0)$ at the event horizon. This point appears to be a degenerate attractor,\footnote{In the original dynamical system parametrized by $\phi$, the point $(u,du/d\phi)=(1,0)$ is singular due to the divergence of the differential equations there. After a suitable time reparametrization that removes the divergence, it can be shown that the point $(u,du/d\phi)=(1,0)$ is a degenerate fixed point (the Jacobian eigenvalues vanish). The asymptotically attracting nature of that fixed point can be checked by solving the trajectories in its vicinity, where we get $\left(1-u\right)\propto \left(du/d\phi\right)^{2(2\kappa+1)}$.} and has a completely different dynamical property from the saddle point at the unstable photon sphere. In fact, its non-hyperbolic nature also distinguishes it from the usual stable light rings, which are typically hyperbolic and stable fixed points. We will come back to this point in Sec.~\ref{Sec:lightring}.

\section{Shadow imaging and inner-shadow signatures}\label{Sec:shadow}

The landmark achievements of EHT, which have recently resolved the apparent silhouettes produced by photon orbits captured or strongly lensed around M87* and Sgr A*, have sparked intense attention in the study of black hole shadows and images within general relativity, extended gravitational theories, and quantum-gravity-inspired models (see e.g. Refs. \cite{Zakharov:2004sg,Mishra2019,Banerjee2020,Chen:2020aix,Vagnozzi2022c,Banerjee2022,Chen:2022lct,Ye2023,Pedrotti2024,Chen:2024ibc,Wang2025,Pantig2024,Ali2024a,Yang2025,Chen:2025jay,Al-Badawi2025,Kala2025,Chen-Yiqian2025,Pantig2025a,Fathi2025,Wang2025b,Tsukamoto:2020bjm,Tsukamoto:2025hbz,Zhen2026}). Black hole shadows provide a powerful observational probe of the strong-field regime and the near-horizon geometry, where deviations from general relativity may leave observable imprints. 

Prompted by these recent developments, in this section we examine how the modified causal and geometric properties of Lorentzian-Euclidean black holes manifest in their shadow images.

After introducing the setup in Sec. \ref{Sec:setup}, the black hole images are explored in Sec. \ref{subsec:shadowimages}, while further details on the inner shadows are given in Sec. \ref{Sec:inner-shadow}.

\subsection{Ray-tracing setup}\label{Sec:setup}

We will carry out the shadow-image simulation using a simple model for the light sources surrounding  Lorentzian-Euclidean black holes: a geometrically and optically thin accretion disk. As matter in the disk gradually spirals inward, its temperature increases and electromagnetic radiation is emitted. Therefore, the entire thin-disk structure can be regarded as a disk-like distribution of emitters whose intensity profile, which we will introduce shortly, depends on the spatial coordinates only through the radial variable $r$. 

The hypothesis of optical thinness entails that photons are not absorbed as they pass through the disk after emission. Consequently, each time photon trajectories intersect the disk, they \textit{effectively accumulate} additional intensity, and if this process occurs repeatedly, the corresponding orbits contribute to higher-order rings on top of the direct emission on the image plane \cite{Gralla:2019xty}. Despite its simplicity, this framework is expected to capture the main image features of  black holes embedded in  more general light-source configurations \cite{Gralla:2019xty,Johnson:2019ljv}. 

We further assume a face-on orientation, motivated by the observational geometry of M87*, where the inclination angle between the disk axis and the observer line of sight is believed to be small \cite{CraigWalker:2018vam}. Using the setup of Fig.~\ref{fig:trajectory}, where the observer is located at $(X,Y)=(100,0)$ in units  $2M=1$, the accretion disk plane is thus represented by the vertical green line along the $X=0$ axis. 

Following the standard radiative transfer relation, the specific intensities $I_{\nu_e}$ and $I_{\nu_o}$, corresponding to frequencies $\nu_e$ and $\nu_o$ in the emitter and observer  frames, respectively, are related by \cite{Lindquist:1966igj}
\begin{equation}
\frac{I_{\nu_e}}{\nu_e^3}=\frac{I_{\nu_o}}{\nu_o^3}\,,\label{IEIO}
\end{equation}
in the absence of absorption. For each photon path $\mathcal{C}$, let $r_k$ denote the radius where the light ray pierces the disk plane. To compute the bolometric observed intensity $I_o$, we proceed as follows: at each piercing $r_k$, we first apply formula   \eqref{IEIO}, take into account the redshift factor $\mathcal{G}:=\nu_o/\nu_e$, and then integrate the specific observed intensity $I_{\nu_o}$ over $\nu_o$ at that piercing.  Repeating this integration at all intersections and summing the contributions yields the total observed intensity $I_o$ for the trajectory. Supposing, for simplicity, that the emitter-frame specific intensity is monochromatic with characteristic frequency $\nu_*$ and depends only on $r$, i.e., 
\begin{equation}
I_{\nu_e}=I_e(r)\delta(\nu_e-\nu_*)\,,
\end{equation}
the observed intensity $I_o$ can be expressed as{\footnote{It is possible to take the factor $\mathcal{G}^3$ rather than $\mathcal{G}^4$ in formula \eqref{I-0-formula-1}. Such an alternative choice allows one to get $I_o$ directly for a broadband emission \cite{Gralla:2020srx}. However, the differences between these two options are negligible and can  degenerate with emission profiles.}}
\begin{equation}\label{I-0-formula-1}
I_o=\sum_{k\in\mathcal{C}}\mathcal{G}(r_k)^4 I_e(r_k)\,.
\end{equation}
For the emission profile $I_e(r)$, we employ the Gralla-Lupsasca-Marrone (GLM) emission model \cite{Gralla:2020srx}:
\begin{equation}\label{GLM}
I_e(r;\gamma,\mu,\sigma)=\frac{\textrm{exp}{\left\{-\frac{1}{2}\left[\gamma+\textrm{arcsinh}\left(\frac{r-\mu}{\sigma}\right)\right]^2\right\}}}{\sqrt{\left(r-\mu\right)^2+\sigma^2}}\,,
\end{equation}
where the three phenomenological parameters $\gamma,\mu$, and $\sigma$ control the profile shape. See also Refs.~\cite{Olmo:2023lil,Rosa:2023qcv,daSilva:2023jxa,DeMartino:2023ovj,Macedo:2024qky,Macedo:2025ipc,Diaz-Guerra:2025jip} where the GLM emission profile was considered in the image modeling of compact objects. 

As shown in Fig.~\ref{fig:trajectory},  tracing photon orbits backward from the observer for impact parameters $b$  smaller than the critical value $b_c$ reveals that these paths never enter the black hole; instead, they rotate around it an infinite number of times and get arbitrarily close to $r=2M$. To identify the image characteristics generated by these light rays, the emission profile should extend down to $r=2M$. We thus adopt the GLM function \eqref{GLM} with $(\gamma,\mu,\sigma)=(-3/2,0,M/2)$. The ensuing emission profile peaks slightly away from the origin ($r\gtrsim0$) and decays to zero as $r\rightarrow\infty$, thereby ensuring nonvanishing intensity near $r\sim2M$.

\subsection{The shadow images}\label{subsec:shadowimages}

Using the setup introduced in the previous section,  we now examine the images of  Lorentzian-Euclidean black holes.

\begin{figure}[htbp!]
\centering
\includegraphics[width=240pt]{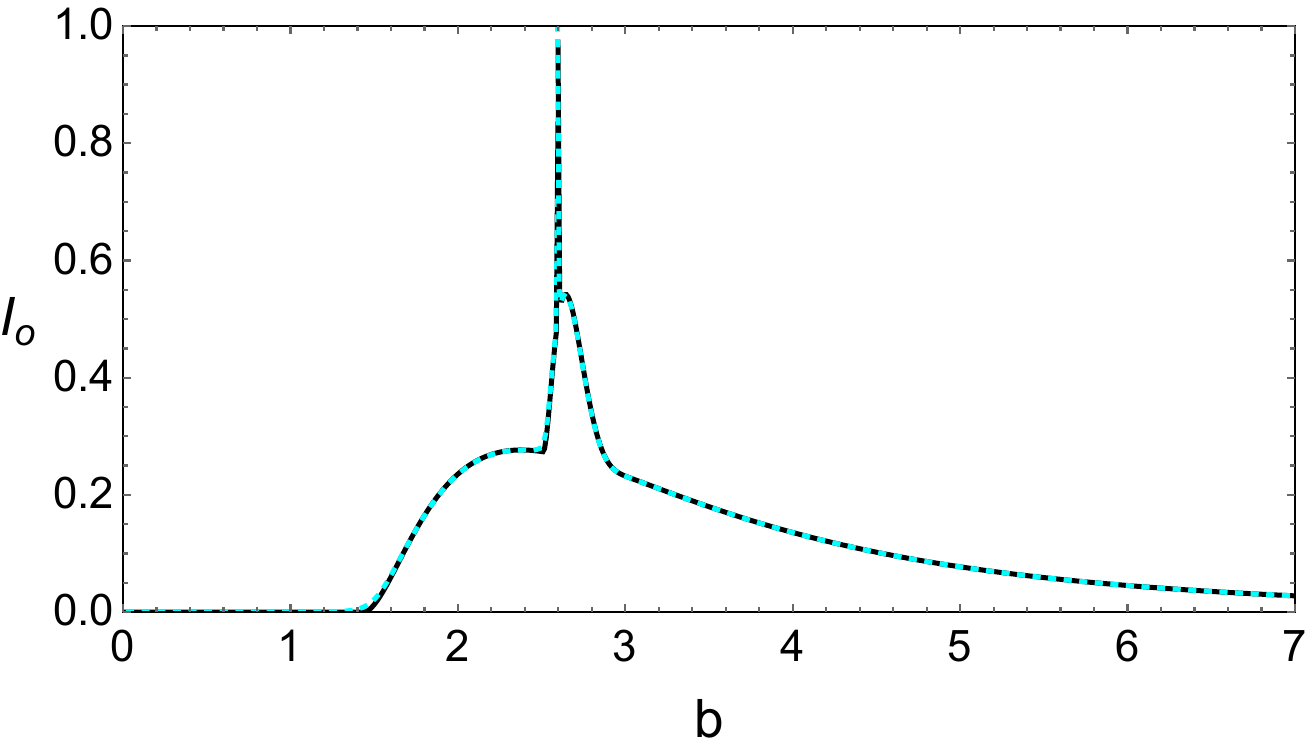}
\caption{Observed intensity $I_o$ as a function of the impact parameter $b$ for the GLM emission profile \eqref{GLM} and with units $2M=1$. The black (resp. cyan dashed) curve represents the Schwarzschild (resp. Lorentzian-Euclidean) black hole. The two curves nearly coincide, with a slight difference appearing only near the inner-shadow boundary around $b\sim1.434$; a zoomed-in view of this region is provided in Fig. \ref{fig:innershadowzoom}. }
\label{fig:Iob}
\end{figure}

\begin{figure}[htbp!]
\centering
\includegraphics[width=120pt]{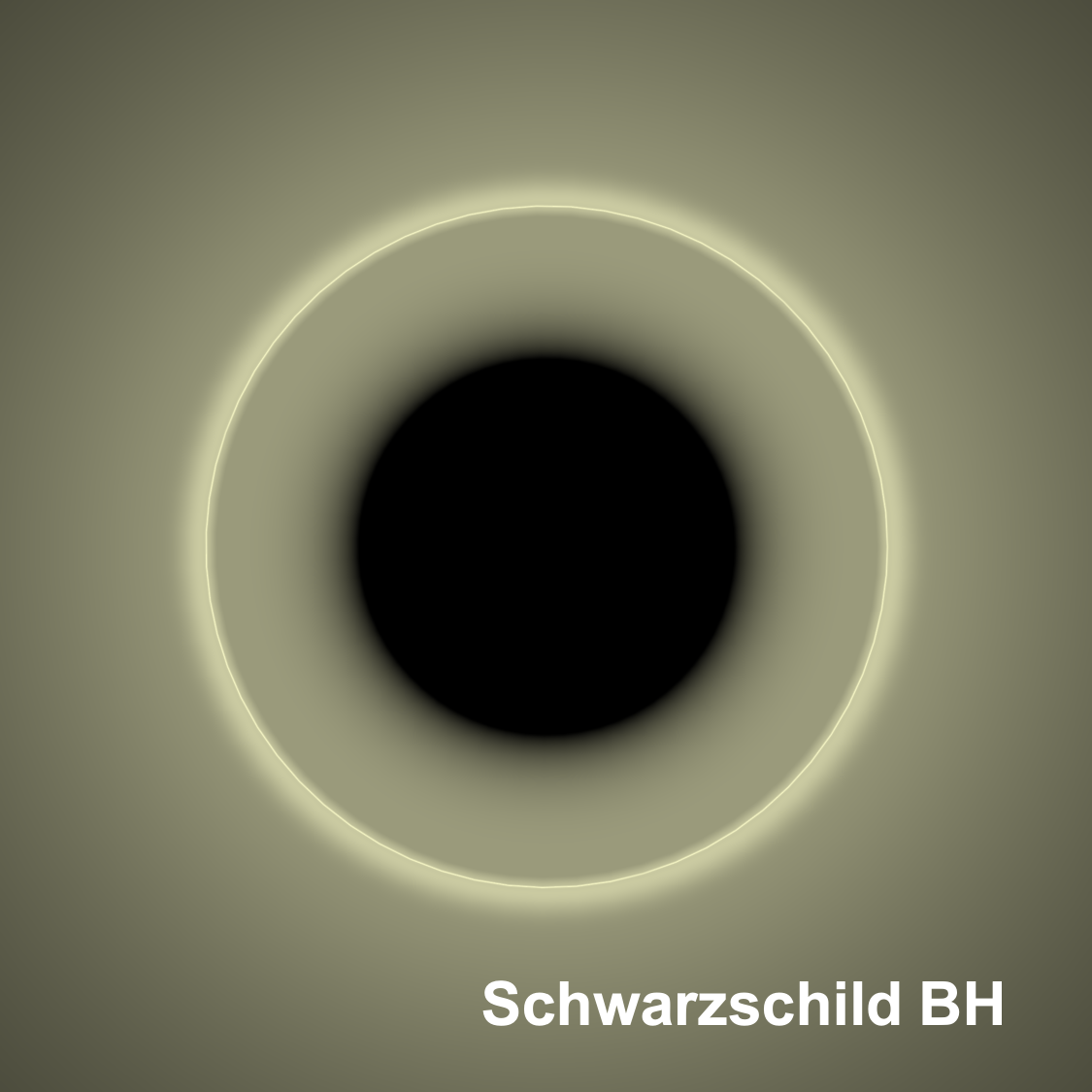}
\includegraphics[width=120pt]{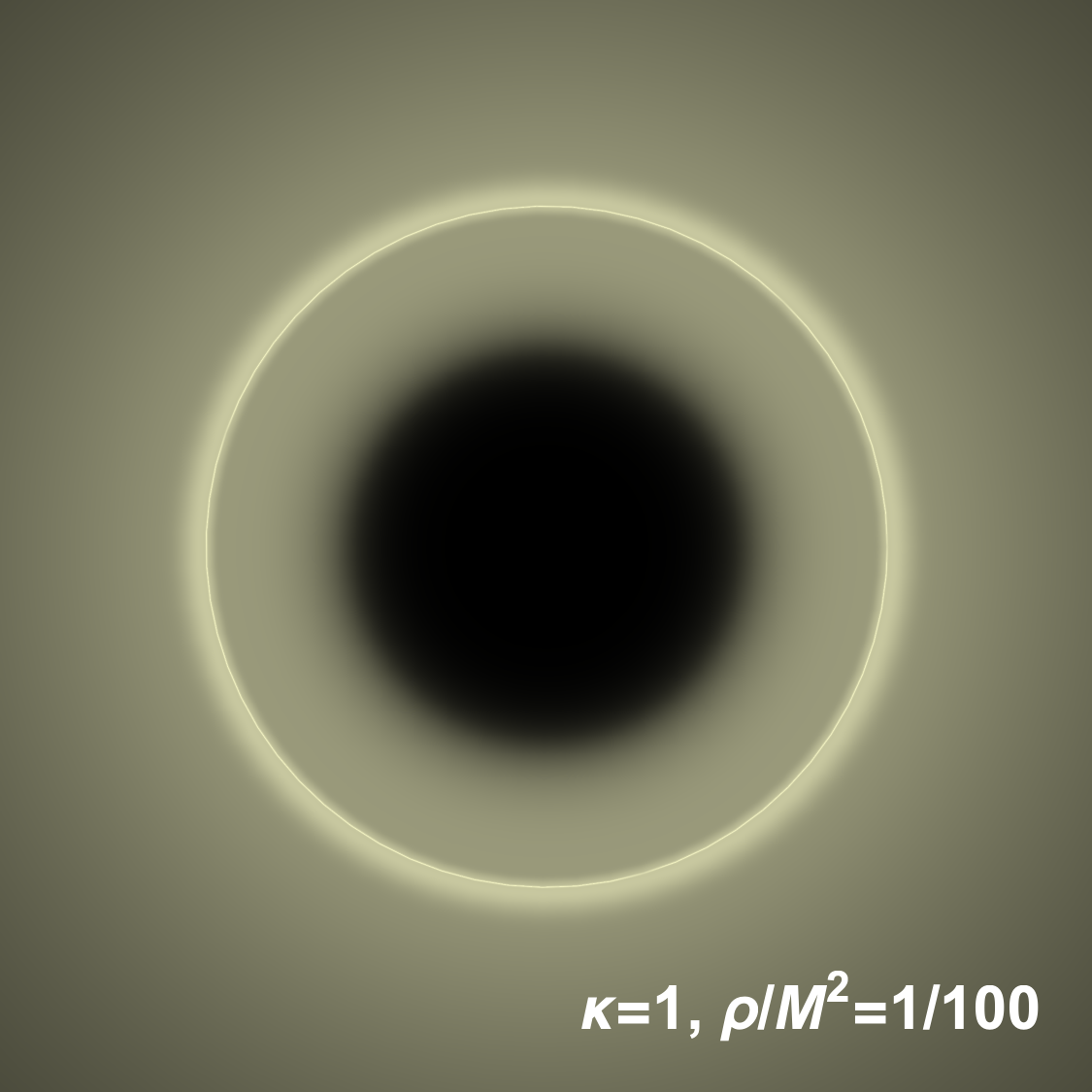}
\caption{Shadow images of the Schwarzschild solution (left) and the Lorentzian-Euclidean black hole with $\kappa=1$ and $\rho/M^2=1/100$ (right). The Lorentzian-Euclidean model produces an inner shadow with a somewhat more diffuse boundary compared to the sharply defined one in the Schwarzschild geometry.}
\label{fig:image}
\end{figure}

In Fig.~\ref{fig:Iob}, we display the observed intensity $I_o$, normalized by its peak value, as a function of the impact parameter $b$ for the GLM emission profile, using units in which $2M=1$. The black curve refers to the Schwarzschild solution, while the cyan dashed one represents the Lorentzian-Euclidean black hole. The two curves nearly overlap and exhibit a sharp peak near  $b\sim\sqrt{27}/2$, associated with higher-order images (see the bright rings in Fig.~\ref{fig:image}) produced by strongly lensed light rays that orbit the black hole multiple times in the vicinity of the photon sphere. Since the geometric modifications introduced by the Lorentzian-Euclidean model relative to the Schwarzschild case are very tiny near the photon sphere,  no significant difference in image features arises in this region. As we illustrate in Fig.~\ref{fig:image}, the overall images of the Schwarzschild (left) and the Lorentzian-Euclidean (right) black holes are thus nearly indistinguishable. 

\begin{figure}[htbp!]
\centering
\includegraphics[width=245pt]{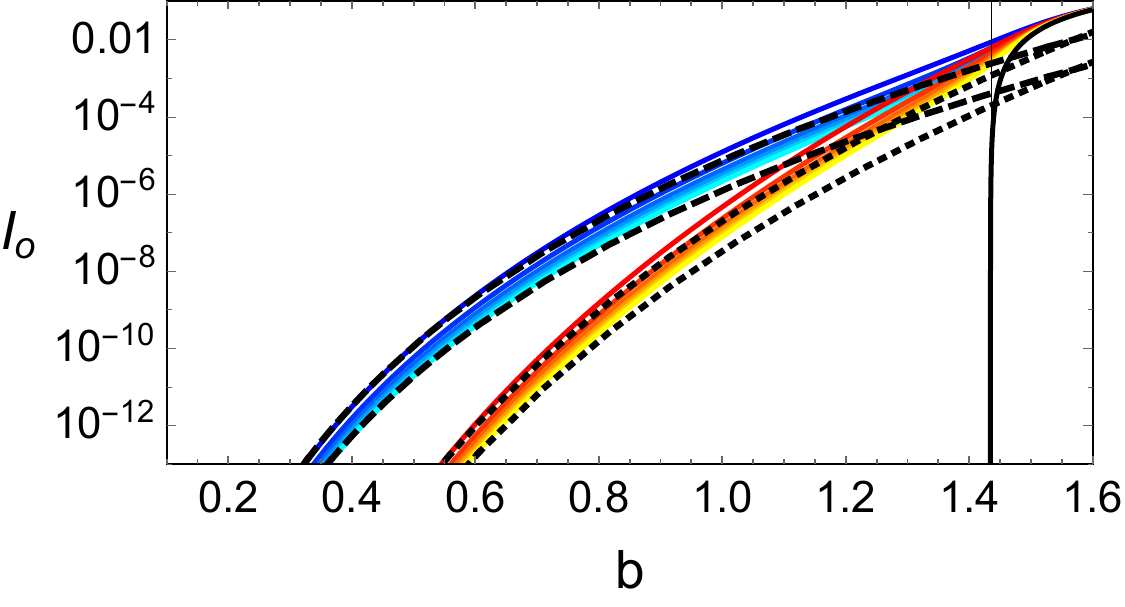}
\caption{Observed intensity $I_o$ in logarithmic scale near the inner shadow, assuming the GLM emission profile \eqref{GLM} and units with $2M=1$. The intensity for the Schwarzschild black hole (black solid curve) falls to zero at the inner-shadow boundary (vertical line at $b=1.434$). Colored curves refer to the Lorentzian-Euclidean black hole, and show an excess intensity. The blueish curves represent $\kappa=1$, while the reddish ones stand for $\kappa=2$. Within each set, the curves from top to bottom correspond to $\rho=1/400$, $1/800$, $1/1200$, $1/1600$, $1/2000$, and $1/2400$, respectively. As $\kappa$ increases or $\rho$ decreases, the excess intensity inside the inner shadow diminishes. The shaded and dotted curves denote the analytic formula \eqref{analyticIo} for $\kappa=1$ and $\kappa=2$, respectively; the upper (resp. lower) curve indicates $\rho=1/400$ (resp. $\rho=1/2400$).}
\label{fig:innershadowzoom}
\end{figure}

Notable features emerge however near the boundary of the so-called inner shadow, i.e., the leftmost tail of the observed intensity $I_o$ occurring at $b\sim1.434$ in Fig.~\ref{fig:Iob}, or equivalently  the boundary of the central dark regions in  Fig.~\ref{fig:image}. This is due to the fact that the Lorentzian-Euclidean black hole exhibits in this area a slightly higher intensity compared to the Schwarzschild solution. This excess is displayed in more detail in Fig.~\ref{fig:innershadowzoom}, where we focus on the inner-shadow region using a logarithmic scale for the observed intensity. 

For the Schwarzschild model (black curve in Fig.~\ref{fig:innershadowzoom}), the intensity drops to zero inside the inner shadow \cite{Chael:2021rjo}. Since the inner-shadow boundary corresponds to the direct emission from the event horizon, photon trajectories, connecting the observer to the horizon, do not intersect the accretion disk; consequently, this region appears completely dark. In contrast,  for the Lorentzian-Euclidean black hole, light rays never cross the event horizon and, in principle, can pierce the accretion disk near $r=2M$ infinitely many times. When tracing the rays backward in time, however, only the first few piercings can contribute appreciably to the observed intensity, as subsequent crossings are heavily suppressed by gravitational redshift. The resulting excess intensity inside the inner shadow arises from these additional piercings, and its dependence on the impact parameter $b$ is controlled by the parameters $\kappa$ and $\rho$ appearing in $\varepsilon(r)$ (cf. Eq. \eqref{epsilonr}). As one can see from the colored curves in Fig.~\ref{fig:innershadowzoom}, the excess intensity inside the inner shadow can be enhanced by either decreasing $\kappa$ or increasing $\rho$.

Returning to Fig.~\ref{fig:image}, we see that the inner shadow of Schwarzschild black holes displays a sharp boundary, whereas that of Lorentzian-Euclidean models appears with a slightly \qm{fuzzy} boundary and is not completely dark.

\subsection{More on the inner shadows}\label{Sec:inner-shadow}

As shown in Fig.~\ref{fig:innershadowzoom}, the excess inner-shadow intensity of  Lorentzian-Euclidean black holes depends on the parameters $\rho$ and $\kappa$ occurring in the function $\varepsilon(r)$. In this section, we derive the analytic expression for this excess near the center of the inner shadow, i.e., in the regime $b/M\ll1$. 

We begin by considering the right-hand side of Eq.~\eqref{angularint}, where we perform an integration from the observer location $r\rightarrow\infty$ down to $r\rightarrow 2M$, noting that the corrections arising from $\varepsilon$ are negligible throughout this domain. For trajectories with small impact parameter $b$, the first term in the square root on the right-hand side of Eq.~\eqref{angularint} dominates, and using  Eq.~\eqref{magictrick} we find
\begin{align}
\Delta(\sqrt{\varepsilon}\phi)\approx-b\int_\infty^{2M}{\frac{dr}{r^2}}=\frac{b}{2M}\,.\label{estimatephi}
\end{align}
Owing to the strong redshift near $r=2M$, the total intensity for each small-$b$  trajectory is primarily due to the first intersection in the backward ray-tracing between the observer and the disk. From Eq.~\eqref{estimatephi}, the radius $r=r_1$ of this crossing satisfies
\begin{equation}
\frac{\pi}{2}\sqrt{\varepsilon(r_1)}=\frac{b}{2M}\,.\label{r1b}
\end{equation}
Since $r_1$ lies extremely close to the horizon, we set $r_1=2M+\delta r$. Substituting Eq.~\eqref{epsilonr} into the above identity determines $\delta r$, and hence $r_1$ reads as
\begin{equation}
r_1\approx 2M+\sqrt{\rho}\left(\frac{b}{\pi M}\right)^{2(2\kappa+1)}\,.
\end{equation}
According to the radiative transfer formula, the observed intensity contributed by the crossing at $r_1$, which is expected to dominate the entire $I_o$, is thus given by 
\begin{align}
    I_o&\approx I_e(2M)(-g_{tt}(r_1))^2\nonumber\\
    &\approx I_e(2M)\varepsilon(r_1)^2\frac{\rho}{4M^2}\left(\frac{b}{\pi M}\right)^{4(2\kappa+1)}\nonumber\\
    &\approx I_e(2M)\left(\frac{b}{\pi M}\right)^{8\kappa+8}\frac{\rho}{4M^2}\,,\label{analyticIo}
\end{align}
where we have used $-g_{tt}=\varepsilon\left(1-2M/r\right)$ (cf. Eq. \eqref{Lorentzian-Euclidean-Schwarzschild}). 

As illustrated in Fig. \ref{fig:innershadowzoom}, the analytic formula \eqref{analyticIo} provides an excellent match to the numerical results in the limit $b/M\ll1$.

\section{Energy accumulation and backreaction effects: differences between the event horizon and stable light rings}\label{Sec:lightring}

A distinctive aspect of the Lorentzian-Euclidean black hole is that it allows photons,  particles, and energy to amass near the horizon. This behavior resembles that of stable light rings that typically appear in models of horizonless compact objects \cite{Cunha:2017qtt,DiFilippo:2024ddg,Chen:2024ibc}. Nevertheless, as we will show, the two phenomena differ significantly in how the backreaction sourced by the accumulated energy modifies the spacetime geometry.

As a simple illustration, we model the build-up of light rays and energy near  $r=2M$ by considering a large number of spherical shells surrounding the Lorentzian-Euclidean black hole at radii just above the horizon. In the continuum-shell limit,  following the procedure outlined in Refs. \cite{DiFilippo:2024poc,Cunha:2025oeu}, the resulting spacetime configuration can be described by the following metric:
\begin{align}
ds^2=&-\psi(r)\left(1-\frac{2m(r)}{r}\right)dt^2+\frac{dr^2}{\left(1-\frac{2m(r)}{r}\right)}
\nonumber \\
&+r^2 \left(d \theta^2  + \sin^2 \theta \; d \phi^2 \right)\,,\label{shellmetric}
\end{align}
where $m(r)$ is a smooth mass function that embodies the distribution of the surrounding shells, and the function $\psi(r)$ depends on $m(r)$ as well as the regularization parameters $\rho$ and $\kappa$. The new surface that collects energy in the geometry \eqref{shellmetric} is determined by the equation
\begin{equation}
C(r):= r-2m(r)=0\,.\label{Cfunction}
\end{equation}
Adopting an analysis similar to that of Ref. \cite{Cunha:2025oeu}, we introduce small perturbations of $m(r)$ through
\begin{equation}
m(r)=m_0(r)+\delta m(r)\,,\label{mex}
\end{equation}
where $m_0(r)$ stands for the unperturbed mass function and $\delta m(r)$ encodes the small contribution from the additional energy residing near the accumulation surface. Then, we suppose that this extra energy  shifts the location of this  surface from $r_\textrm{as}^0$ to 
\begin{equation}
r_\textrm{as}=r_\textrm{as}^0+\delta r_\textrm{as}\,.\label{rasex}
\end{equation}
The relation between $\delta r_\textrm{as}$ and $\delta m(r)$ can be obtained by plugging Eqs.~\eqref{mex} and \eqref{rasex} into Eq.~\eqref{Cfunction}, which upon expanding  to first order, yields
\begin{equation}
\delta r_\textrm{as}\approx\frac{2\delta m(r_\textrm{as})}{C'(r_\textrm{as})}\,.
\end{equation}
Therefore, for a positive $\delta m(r_\textrm{as})$ due to the energy build-up, whether the aggregation surface shrinks or expands depends on the sign of $C'(r_\textrm{as})$. Since $C(\infty)>0$ (i.e., $C(r)$ attains positive values at spatial infinity),  $C(r_\textrm{as})=0$, and assuming that the function $C(r)$ has no other roots for $r>r_\textrm{as}$ so that $r_\textrm{as}$ represents the outermost horizon, we obtain $C'(r_\textrm{as})>0$. As a result, the accumulation surface expands under the backreaction induced by the additional energy. This behavior differs from that of typical stable light rings, which would shrink as energy builds up around them, as shown in Ref. \cite{Cunha:2025oeu}.

The existence of stable light rings has been conjectured to trigger spacetime instabilities through the continuous build-up of energy \cite{Keir:2014oka,Cardoso:2014sna,Cunha:2022gde}. Importantly, Ref.~\cite{Cunha:2025oeu} did not prove that stable light rings always lead to instabilities, but instead examined how small perturbations sourced by gathered energy modify the underlying spacetime. In the same spirit, our analysis shows only that, at a perturbative level, the backreaction on the Lorentzian-Euclidean Schwarzschild geometry is distinct from that associated with standard stable light rings, despite the fact that the event horizon $r=2M$ in our model allows for energy concentration. This difference does not imply that Lorentzian-Euclidean Schwarzschild black holes are free from such instability issues. Whether the energy accumulation in our configuration triggers spacetime instabilities remains an open question to be explored elsewhere.

\section{Discussion and conclusions}
\label{Sec:Conclusion}

The Lorentzian-Euclidean model is characterized by a metric that undergoes a signature change at $r=2M$, with the geometry sharing similar properties with the Euclidean Schwarzschild solution for $r<2M$. A remarkable facet is that photons never cross the event horizon, but they are bent into trajectories that can circle near $r=2M$ infinitely many times. 

The main aim of this paper has been to investigate the shadow images produced by Lorentzian-Euclidean black holes and to identify distinctive aspects associated with this novel behavior of light rays.

A key outcome of our analysis is the presence of an excess intensity of the image in the inner shadow region. In the standard Schwarzschild scenario, photons contributing to the inner shadow connect the observer and the event horizon without intersecting any emission sources along their path, and hence the inner shadow appears completely dark. On the other hand, the inner shadow featuring the Lorentzian-Euclidean black hole may not be entirely dark owing to the highly redshifted emissions originating very close to $r=2M$. The analytic treatment presented in Sec.~\ref{Sec:inner-shadow}, which matches closely the numerical results (see Fig.~\ref{fig:innershadowzoom}), shows that this excess intensity can be described in terms of the regularizing function $\varepsilon(r)$ and its parameters (cf. Eq. \eqref{epsilonr}), which represent an intrinsic feature of our model required to remove pathological $\delta$-contributions in the spacetime.

We emphasize that the excess intensity in the inner shadow of the images, cast by compact objects, can potentially be a generic aspect of horizon-scale modifications to black hole geometry. In the Lorentzian-Euclidean framework, this distinctive image characteristic arises from the signature shift at $r=2M$, which alters the horizon properties and the particle dynamics in its vicinity. Albeit not strictly necessary, one can interpret this signature change as a phenomenological consequence of some quantum effects, a perspective that is theoretically motivated and widely adopted in quantum cosmology \cite{Gibbons:1990ns,Hartle:1983ai} and, more recently, in black hole physics \cite{Bojowald:2016itl,Bojowald:2018xxu,Tzanavaris:2024acr}. In models of horizonless compact bodies where event horizons never form due to quantum effects, even if strong redshifts significantly darken additional light rings in the inner shadow, the excess intensity still remains a robust image characteristic  \cite{Bacchini:2021fig,Chen:2024ibc}. Therefore, if the signature-varying property of  Lorentzian-Euclidean black holes has a quantum origin, this paper provides another explicit example supporting the possibility of probing horizon-scale quantum effects through the excess inner-shadow intensity, assuming the existence of light emission near the horizon for black holes or of the would-be-horizon for horizonless compact objects.

Another crucial result of this paper is the following. Although infalling light rays, accreting matter, and the associated energy can be gathered at the event horizon of Lorentzian-Euclidean black holes, our analysis of  Sec.~\ref{Sec:lightring} demonstrates that its response to the backreaction from the piled-up energy differs from that of stable light rings in typical horizonless compact object models. More explicitly, using a perturbative approach, we find that the event horizon of the Lorentzian-Euclidean spacetime expands as a geometric response to the energy collection, while stable light rings generically shrink under similar backreaction effects \cite{Cunha:2025oeu}. This expansion behavior aligns with the investigation of Ref.~\cite{Capozziello:2025wwl}, which shows that the matter accretion onto  Lorentzian-Euclidean black holes also increases their mass.

A final remark has to be put forward. The inner shadow characteristics studied in Sec. \ref{Sec:shadow}, the accumulation process addressed in Sec.~\ref{Sec:lightring}, as well as the matter accretion mechanism examined in Ref.~\cite{Capozziello:2025wwl}, are analyzed under the assumption that the backreaction of particles building up near $r=2M$ is either neglected or treated only at a perturbative level. An important point to explore is how such amassing might modify the entire geometry non-perturbatively, thereby potentially generating additional observational signatures. We leave this direction of research to future work.  

\section*{Acknowledgments}
EB and SC thank Silvia De Bianchi for useful discussions on the {\it atemporality}.
EB and SC acknowledge the support of  {\it Istituto Nazionale di Fisica Nucleare} (INFN), {\it Iniziative Specifiche}  MOONLIGHT2.  SC acknowledges the support of  {\it Istituto Nazionale di Fisica Nucleare} (INFN), {\it Iniziative Specifiche}   QGSKY. SC is grateful to the {\it Gruppo Nazionale di Fisica Matematica} (GNFM) of {\it Istituto Nazionale di Alta Matematica} (INDAM) for support. CYC is supported by the Special Postdoctoral Researcher (SPDR) Program at RIKEN and RIKEN Incentive Research Grant (Shoreikadai) 2025.

\bibliography{references}

\end{document}